\documentclass[%
 reprint,
 amsmath,amssymb,
 aps,
 pre,
]{revtex4-2}

\usepackage{graphicx}
\usepackage{dcolumn}
\usepackage{bm}
\usepackage{xcolor}

\begin{document}

\title{A Statistical-Mechanical Model for Dipolar Chain Formation}

\author{Zhongqi Liang}
\email{zhongqi.liang@stonybrook.edu}
\author{Jes\'us Per\'ez-R\'ios}
\affiliation{Department of Physics and Astronomy, Stony Brook University, Stony Brook, New York, 11790, USA}

\date{\today}

\begin{abstract}
Dipolar fluids are known to exhibit complex self-assembly at low temperatures, yet a compact thermodynamic description of their aggregate statistics has remained elusive. Using molecular dynamics simulations of Stockmayer particles with a purely repulsive WCA core, we confirm that over broad regions of the ($\rho$, $T$) phase space the chain-size distribution follows an exponential decay with characteristic size $s_0$. Within this regime, we find that $s_0$ can be accurately described by an effective free energy $\phi$ that incorporates translational entropy, bonding energy, and a crowding penalty. Identifying deviations from this ideal scaling provides a further division of the phase space into four regions. Therefore, our results locate a regime of relatively simple chain statistics and offer an alternative regime-based perspective on dipolar self-assembly.

\end{abstract}

\maketitle

    The phase behavior of interacting self-assembling dipoles has been the subject of theoretical and computational debate for decades without a definitive resolution~\cite{Caillol1993,Weis1993,Levesque1994,Sear1996,vanRoij1996,Camp2000,Teixeira2000,Hynninen2005,Ganzenmuller2009,Rovigatti2011}. At sufficiently high temperatures, the thermal motion renders the fluid nearly isotropic. However, as the temperature drops, the anisotropic dipole-dipole interaction drives the dipoles to self-assemble into chains and more complex network-like structures before a traditional liquid-vapor coexistence line can be identified~\cite{Sear1996,vanRoij1996}. In an alternative picture, Tlusty and Safran proposed that a topological phase transition can occur between an end-rich gas phase and a junction-rich liquid phase, where an end connects only to one neighbor, but a junction connects to three or more~\cite{TS2000}. However, this scenario is complicated by the appearance of another competing topological structure, namely rings formed by chain closure, and a coexistence line has yet to be found~\cite{Rovigatti2011}. Most early work focuses on a hard-sphere model for dipoles ~\cite{Caillol1993,Weis1993,Camp2000,Rovigatti2011}. More recently, Yukawa hard spheres~\cite{Ganzenmuller2009, Scalise2009}, soft spheres~\cite{Hynninen2005, Jia2010}, Stockmayer particles~\cite{Hentschke2007, Bartke2007,Shock2020, Liu2024, Staubach2025}, and dumbbells~\cite{Ganzenmuller2007,Liu2024} have been explored as alternative models of dipoles. Nevertheless, no conclusive evidence of phase coexistence has been observed in direct simulations when no additional attractive term is introduced to the potential.

    Such self-assembly into complex topological networks is not unique to dipolar fluids. A similar aggregation process also governs the phase space behavior of patchy particles~\cite{Zhang2004,Bianchi2006,Russo2011}, worm-like surfactant micelles~\cite{Cates1990,Bergstrom2015,Chauhan2024}, colloidal gels~\cite{Zaccarelli2007}, and living polymers~\cite{Cates1987,Likos2001,Dudowicz2004}. Therefore, understanding the thermodynamic behavior of dipolar self-assembly provides critical theoretical insights into the study of associating fluids across various contexts of soft matter, physical chemistry, and materials science. 

    Despite a prolonged search for a phase transition in dipolar systems and the prevalence of self-assembly across many physical contexts, a coarse-grained statistical-mechanical description of dipolar self-assembly based on simulation data remains missing. In this letter, we attempt to bridge this gap by introducing a compact coarse-grained statistical-mechanical description for dipolar chain formation in a Stockmayer fluid with purely repulsive cores. With data from extensive molecular dynamics (MD) simulations, we confirm that the chain size distribution follows an exponential decay in a broad sector of the $(\rho, T)$ phase space, as noted before in Ref.~\cite{deGennes1970,Jordan1973,Tavares1999,Teixeira2000,Rovigatti2011}. Further, we show that, in this regime, the emergent characteristic chain size $s_0$ is governed by an effective free energy $\phi$ that accounts for bonding, crowding, and translational entropy. By systematically tracking deviations from this ideal scaling, we are able to produce a regime map of the sampled phase space separated into four regions, offering an alternative regime-based perspective on dipolar self-assembly.

    We perform MD simulations of $N=3000$ Stockmayer particles with repulsive cores placed in a cubic box with periodic boundary conditions using LAMMPS~\cite{LAMMPS}. This system size provides sufficient statistical sampling of the relevant aggregate populations while remaining computationally tractable for the extensive exploration of the $(\rho,T)$ phase space. Additional simulations from $N=2000$ to $5000$ reveal no statistically significant systematic dependence of $s_0$ on $N$ at any tested state point~\footnote{Additional simulations are performed at $\mu=1.80$, $\rho=0.0054, 0.027$, using $N=2000$ to $5000$ and six temperatures spanning $T\in[0.24, 0.48]$. No statistically significant systematic dependence of $s_0$ on $N$ is observed. In the exponential regime, the relative standard deviation across system sizes ranges from 0.4\% to 6.4\%. At $T=0.24$, the variations are larger but $s_0$ is not well-defined since $n_c(s)$ is nonexponential.}. The temperature is controlled through a Nos\'e-Hoover thermostat to maintain the system in a canonical $NVT$ ensemble. The inter-particle interaction is characterized by a Weeks-Chandler-Andersen (WCA) repulsion at short-range
    \begin{equation} \label{eqn:wca}
        V_{ij}^{WCA} (r_{ij}) =
        \begin{cases}
        4\epsilon [(\frac{\sigma}{r_{ij}})^{12} - (\frac{\sigma}{r_{ij}})^6]+\epsilon & r_{ij} \le 2^{1/6} \sigma \\
            0   & r_{ij} > 2^{1/6} \sigma \\
        \end{cases}
    \end{equation}
    where $\epsilon$ and $\sigma$ are the characteristic energy and distance scales of the system and $r_{ij}$ the distance between the $i$th and the $j$th dipole, and a long-range anisotropic dipole-dipole interaction
    \begin{equation} \label{eqn:dd}
        V_{ij}^{dd}(\boldsymbol{r}_{ij})=\frac{1}{4 \pi \epsilon_0 r_{ij}^3}(\boldsymbol{\mu}_i \cdot \boldsymbol{\mu}_j - 3 (\boldsymbol{\mu}_i \cdot \boldsymbol{\hat{r}}_{ij})(\boldsymbol{\mu}_j \cdot \boldsymbol{\hat{r}}_{ij}))
    \end{equation}
    where $\boldsymbol{\mu}_i$ is the dipole moment vector of the $i$th dipole and $\epsilon_0$ the permittivity of free space~\cite{Ivanov2008,Liang2021,Shock2020}~\footnote{Long-range dipole–dipole interactions are evaluated using the LAMMPS \texttt{pppm/dipole} solver. The real-space contribution is truncated at $r_{ij} = 12\sigma$, and the reciprocal-space accuracy is set to $10^{-3}$}.
    
    \begin{figure*}[ht]
        \centering
        \includegraphics[width=0.9\linewidth]{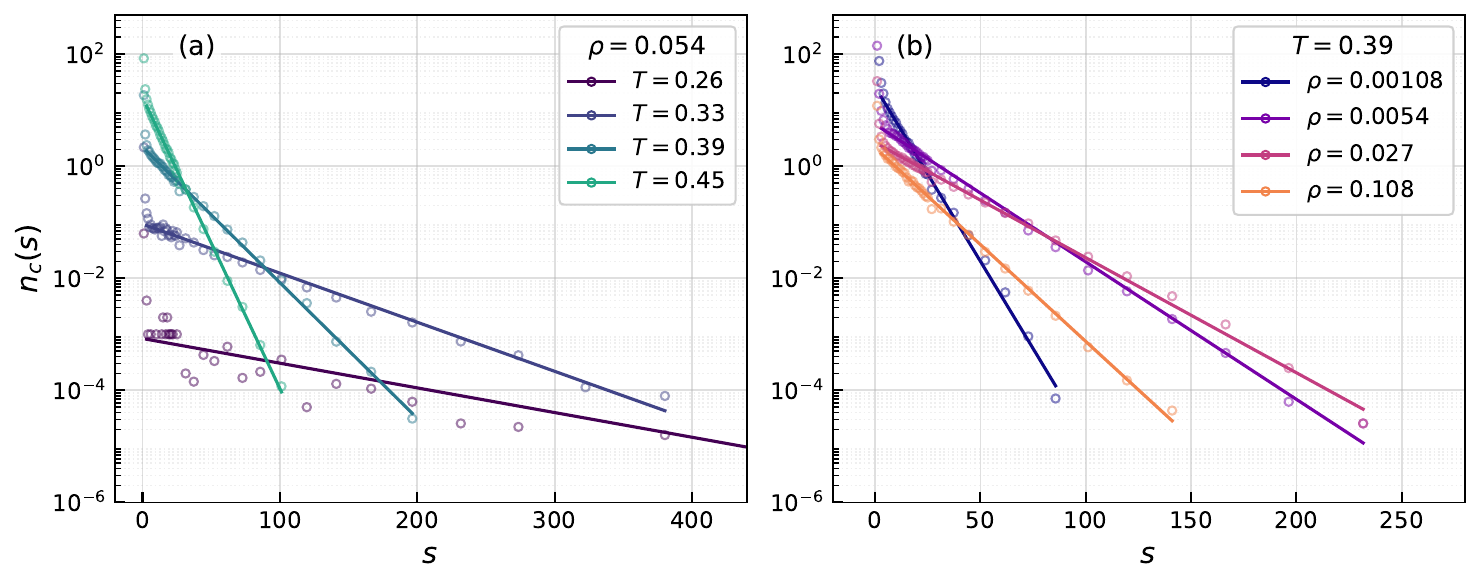}
        \caption{Exponential fits of the number of chains $n_c(s)$ as a function of their size $s$ at $\mu=1.80$. In panel (a), $\rho$ is fixed at 0.054 and $T$ is varied, whereas the $T$ is kept at 0.39 and $\rho$ is varied in panel (b). The hollow circles represent the average values calculated from 1001 freeze frames at each configuration. The fits start from $s=3$ and are shown in solid lines.}
        \label{fig:1}
    \end{figure*}
    
    Simulations are performed in reduced Lennard-Jones (LJ) units where $\epsilon$, $\sigma$, dipolar mass $m$, and the Boltzmann constant $k_B$ are all set to unity. Consequently, physical quantities such as $\rho^* = N\sigma^3/V, T^*= k_B T/\epsilon$ and $\mu^* = \mu/\sqrt{4 \pi \epsilon_0 \sigma^3 \epsilon}$ are scaled accordingly with time measured in units of $\tau^* = \sigma \sqrt{m/\epsilon}$. Hereafter, the asterisk is dropped but assumed, as all quantities are expressed in reduced units. Three dipole strengths of $\mu = 1.55, 1.80, 2.10$ are explored over densities $0.00108 \leq \rho \leq 0.108$ and $\mu$-dependent temperatures, most of which fall within the range of $0.07 \lesssim T/\mu^2 \lesssim 0.16$, overlapping with the chain-forming regime suggested for dipolar hard spheres (DHS) in Refs.~\cite{Weis1993,Levesque1994,Rovigatti2011}~{\footnote{The 14 sampled densities are $\rho=$ 0.00108, 0.00216, 0.0027, 0.0043, 0.0054, 0.0068, 0.0081, 0.0094, 0.0108, 0.018, 0.027, 0.054, 0.081, and 0.108. The temperature ranges are $0.14 \leq T \leq 0.39$, $0.20 \leq T \leq 0.54$, and $0.30 \leq T \leq 0.75$ for $\mu=$ 1.55, 1.80, and 2.10, respectively. Temperatures are sampled more finely at lower $T$, with spacings ranging from 0.02 to 0.05 and larger values used at higher $T$ and $\mu$.}}.

    For each combination of $(\mu, \rho, T)$, the dipolar positions and orientations are first randomly initialized and relaxed to a more stable configuration~\footnote{For relaxation, we run $5 \times 10^4$ time steps of $\Delta t=0.0002$, corresponding to 0.1 - 0.2 fs in real units for molecules such as water.}. Then, after reaching equilibrium, the simulation is continued to collect production data for analysis~\footnote{For equilibration and production, $\Delta t$ is increased to 0.002 for $10^7$ steps to reach equilibrium, and additional $5 \times 10^6$ steps are run to produce sampling data at 1001 freeze frames.}. In each freeze frame, we build bonds between neighboring dipoles $i$ and $j$ with both a spatial ($r_{ij}<1.3)$~\cite{Rovigatti2013} and an energetic criterion ($E_{ij} \leq 1 - 0.8\mu^2$)~\footnote{This specific cutoff is empirically chosen to incorporate a $+1$ positive energy shift from WCA repulsion while finding a balance between preserving chains at low densities and breaking up clustered structures at high densities.}. By analyzing the network topology of these connected substructures, we classify them into chains, rings, or defect clusters~\footnote{The topological classification employs two metrics: the maximum number of bonds connected to a single particle $k$ and the number of closed loops $N_l$. If $k>2$ or $N_l>1$, then it is counted as a defect cluster. $N_l=1$ corresponds to rings and $N_l=0$ indicates the presence of chains.}. In this work, we focus on the analysis of the number of chains $n_c$ as a function of size $s$.

    Despite the structural complexity of individual configurations, the chain-size distribution $n_c(s)$ for $s \geq 3$, excluding monomers and dimers, is well described by an exponential decay over a broad sector of the phase space, as shown in Fig.~\ref{fig:1}. This is consistent with a Boltzmann weight $P_c(s) \propto \exp(- \beta s \Delta F_c)$ for chains, where $P_c(s)$ is the probability of finding a chain of size $s$, $\beta \equiv 1/(k_B T)$, and $\Delta F_c$ represents the effective free-energy cost of adding a monomer to an existing chain. Equivalently, in this simplest picture, one may write $n_c(s) \propto \exp (-s/s_0)$, defining a characteristic chain size $s_0 = (\beta \Delta F_c)^{-1}$.
    
    However, noticeable deviations from this rule can occur at low temperatures, as evidenced by Fig.~\ref{fig:1}(a), when rings and defect clusters become more prominent, pointing to a regime where $\Delta F_c$ can no longer be treated as independent of $s$. We classify state points as belonging to the exponential regime by requiring the linear fit of $\ln(n_c)$ versus $s$ to contain at least 5 data points satisfying $s\geq3$ and produce $R^2 > 0.9$. 

    \begin{figure*}[ht]
        \centering
        \includegraphics[width=0.9\linewidth]{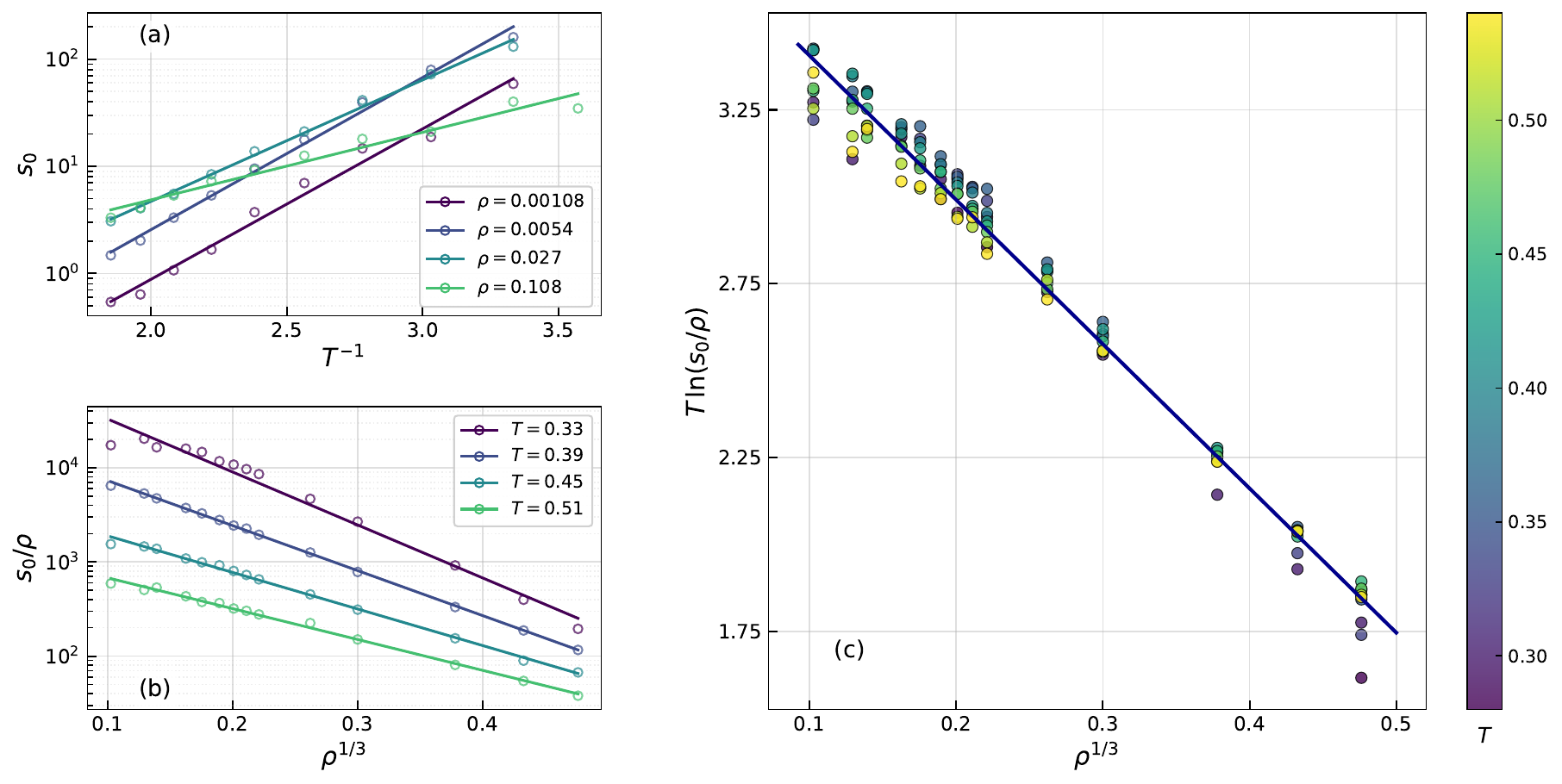}
        \caption{Thermodynamic structure of characteristic chain size $s_0$ as a function of $\rho$ and $T$ at $\mu=1.80$. Uniform binning is used for $s \leq 25$ and exponential binning at larger $s$. Panel (a) shows $s_0 (T)$ at different densities. Panel (b) plots $s_0/\rho$ versus $\rho^{1/3}$ at different values of $T$ to show the dependence of $s_0$ on $\rho$. Then in panel (c), $T \ln(s_0/\rho)$ is plotted against $\rho^{1/3}$ to reveal the underlying structure of $s_0(\rho, T)$. $s_0$ obtained from simulation data are represented by circles while the solid lines denote the linear fits. Colors in panel (c) indicate temperatures.}
        \label{fig:2}
    \end{figure*}

    In the exponential regime, the size-independent monomer-addition free energy $\Delta F_c$ makes $s_0$ a meaningful measure of the extent of chain growth. This parallels equilibrium polymers and worm-like micelles, where reversible scission and recombination lead to an exponential chain-length distribution with a characteristic scale set by the balance between growth and breaking processes~\cite{Cates1987,Cates1990}. Therefore, we interpret $s_0$ as a global equilibrium chain-growth scale for dipolar chains and define
    \begin{equation}
        \phi \equiv -k_B T \ln s_0,
    \end{equation}
    as an effective free energy associated with the overall extent of reversible aggregation.

    There are three physically motivated terms that can contribute to $\phi$. First, the availability of monomers gives the usual mass-action contribution, originating from translational entropy and represented by $-k_B T \ln \rho$. Second, monomer attachment to a chain end is energetically favored by head-to-tail alignment, which produces a $\mu$-dependent characteristic bonding energy $-E(\mu)$. Third, the chain growth process takes place in a dipolar fluid through which monomers move to become attached. Therefore, increasing density also introduces a crowding penalty. The natural geometric length scale associated with the surrounding medium is the mean interparticle spacing $r_m \sim \rho^{-1/3}$. The leading-order density correction controlled by this length scale is therefore proportional to $1/r_m \sim \rho^{1/3}$, giving a positive contribution $C(\mu)\rho^{1/3}$, where $C(\mu)$ is a positive number dependent on $\mu$. This term is sufficient to account for the simulation data over the parameter range considered and is also analogous to the geometric rate control in diffusion-controlled reactions~\cite{North1966}. Combining these terms thus leads to 
    \begin{equation} \label{eqn:hypothesis_phi}
        \phi(\rho, T, \mu) = -k_B T \ln \rho - E(\mu) + C(\mu)\rho^{1/3}
    \end{equation}
    and equivalently
    \begin{equation} \label{eqn:hypothesis_s0}
       s_0(\rho, T, \mu) = \rho \exp(\frac{E(\mu) - C(\mu)\rho^{1/3}}{k_B T}).
    \end{equation}

    This form makes several testable predictions. At fixed $\mu = 1.80$, $E$ and $C$ are constants. In this scenario, when $\rho$ is fixed, $s_0$ should vary exponentially with $1/T$, as confirmed by Fig.~\ref{fig:2}(a). The competition between the translational entropy and the crowding penalty leads to a non-monotonic dependence of $s_0$ on $\rho$ at fixed $T$, consistent with Fig.~\ref{fig:1}(b). Finally, the same expression predicts a linear relation in reduced units
    \begin{equation}
    T\ln(s_0/\rho)=E(\mu)-C(\mu)\rho^{1/3},
    \end{equation}
    which is tested in Fig.~\ref{fig:2}(c). The linear fit gives $R^2=0.9834$, despite a few outliers at low temperatures where rings and defect clusters become significant and the one-scale chain description begins to break down.

    \begin{figure}[t]
        \centering
        \includegraphics[width=0.95\linewidth]{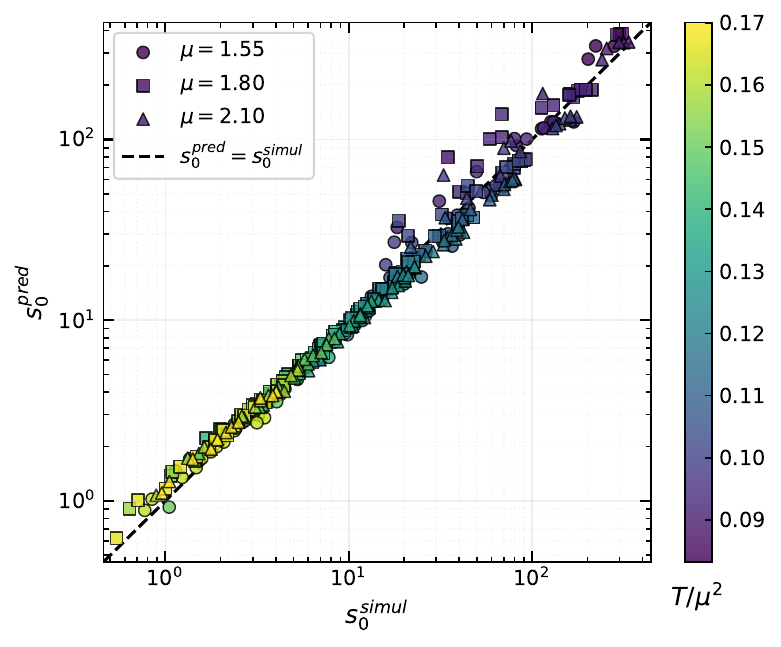}
        \caption{Comparison between $s_0^{pred}$ predicted by the effective free energy $\phi$ and $s_0^{simul}$ calculated from simulation data across three different values of $\mu$. Each point represents a unique $(\rho, T, \mu)$ combination. Colors correspond to normalized $T$ against $\mu^2$. Shapes of scattered points represent different values of $\mu$. Density is not separately encoded.}
        \label{fig:3}
    \end{figure}

    To extend the fixed-$\mu$ description to all dipole strengths, we hypothesize that
    \begin{equation} \label{eqn:hypothesis_E_C}
        E(\mu) = A_1 \mu^2 + B_1 \hspace{1cm} C(\mu) = A_2 \mu^2.
    \end{equation}
    The term for bonding energy, $E(\mu)$, contains a quadratic contribution in $\mu^2$ from the dipole-dipole interaction and a term proportional to $\epsilon=1$ from the WCA repulsion. In contrast, $C(\mu)$ scales the crowding penalty associated with $\rho^{1/3} \sim 1/r_m$. Since $r_m$ is greater than $r_c= 2^{1/6}$ for the WCA potential to be activated in all our simulations, a constant term is not included in $C(\mu)$.
    
    
    Substituting them into Eq.~(\ref{eqn:hypothesis_s0}) produces a global expression for $s_0(\rho, T, \mu)$ that is tested in Fig.~\ref{fig:3}. It shows that the model predicts $s_0$ with good quantitative agreement across different values of $\mu$ except when $s_0$ becomes too small or the system is at relatively low temperatures. Nevertheless, even when all 352 state points that belong to the exponential regime are retained, the line $\ln(s_0^{pred}) = \ln(s_0^{simul})$ yields $R^2 = 0.9864$ after a first fitting of $A_1$, $B_1$, and $A_2$. Since the dipolar interaction energy scales as $\mu^2$, the visualizations are scaled in units of $T/\mu^2$ when different values of $\mu$ are involved.

    \begin{figure}[t!]
        \centering
        \includegraphics[width=0.95\linewidth]{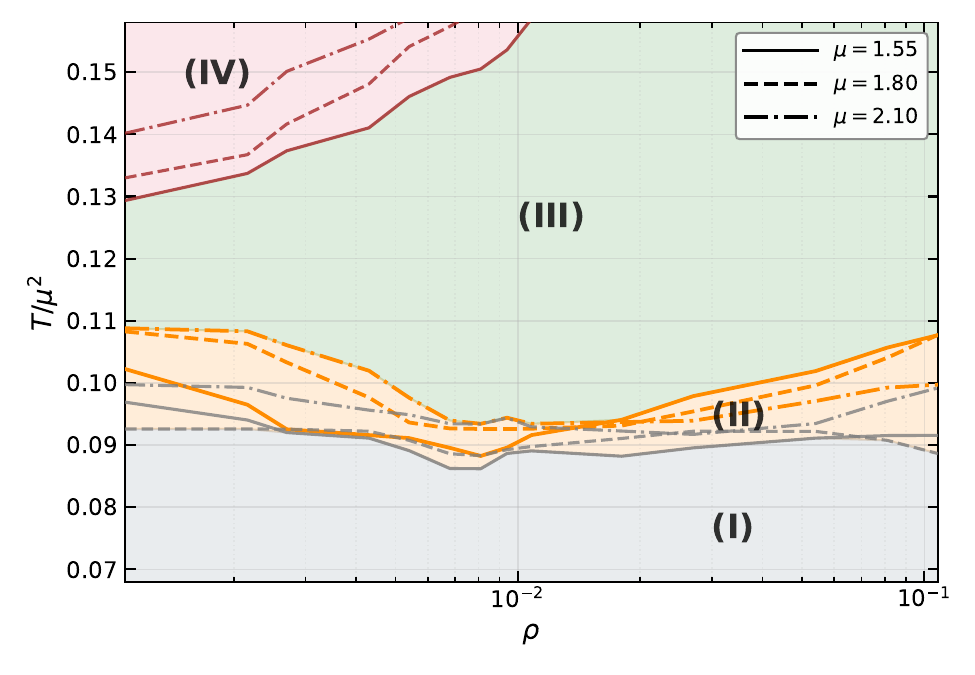}
        \caption{Identification of four different regions of phase space characterized by $\rho$ and $T/\mu^2$ for different values of $\mu$. Region (I): non-exponential regime. Region (II): transitional region where $n_c(s)$ is exponential but $|\ln(s_0^{simul})-\ln(s_0^{pred})|>0.25$. Region (III): thermodynamic chain regime where $s_0$ is accurately predicted from $\phi$ except a few isolated instances. Region (IV): $s_0<2$ at high temperatures and low densities. The boundaries are approximate since only discrete values of $(\rho, T)$ combinations are sampled and marked in different styles for different values of $\mu$.}
        \label{fig:4}
    \end{figure}

    For a more reliable estimate of the parameters, we first exclude the state points where $s_0 < 2$, which indicates the dominance of monomers and dimers over larger aggregates. Then, we implement an iterative filtering procedure that removes, after each fit, any state point with $|\ln(s_0^{simul})-\ln(s_0^{pred})|>0.25$, corresponding to an error in $\phi$ greater than 0.25 $k_B T$. 
    
    This procedure partitions the phase space into four regions, as shown in Fig.~\ref{fig:4}. At low temperatures, with a weak dependence on density, Region (I) corresponds to a highly associative state where open chains are dominated by closed rings and more connected defect clusters, rendering the exponential description of $n_c(s)$ invalid. As $T$ increases, a transitional Region (II) emerges. Here, structures with higher connectivity start to break apart and chains acquire a more significant presence. Although the exponential dependence of $n_c(s)$ starts to appear, $s_0$ can still significantly deviate from the proposed ansatz. However, in Region (III), except for a few isolated points, $s_0$ can be accurately predicted from the effective free energy $\phi$ with $A_1 = 1.336 \pm 0.004$, $B_1 = -0.439 \pm 0.009$, $A_2 = 1.309 \pm 0.007$ and $R^2 = 0.9939$ for 280 combinations of $(\rho, \mu, T)$. This region shows the characteristic exponential chain formation described above, where chain growth is governed by the combined effects of bonding, crowding, and translational entropy. Finally, at high temperatures and low densities, Region (IV) corresponds to a more isotropic and entropy-dominated structure. Thermal fluctuations overwhelm the dipolar bonding energy in a less crowded medium, producing a system rich in monomers and dimers but lacking in extended structures.
    
    As the $\mu$-independent constant term in $E(\mu)$ is naturally associated with short-range WCA repulsion, our results suggest that a system of dipolar soft spheres (DSS) need not be thermodynamically equivalent to that of DHS, especially at small $\mu$. Nevertheless, after replacing the soft WCA core with a hard sphere, we expect that a related thermodynamic description will emerge with $B_1 = 0$ and thus fully restore the strict scaling with $T/\mu^2$, rather than with $T$ and $\mu$ separately. However, verifying this conjecture requires direct simulations of DHS fluids and lies beyond the scope of this work.

    In summary, by studying the chain size distribution $n_c(s)$ in a purely repulsive Stockmayer fluid, we have confirmed that it follows an exponential form over a broad sector of the $(\rho, T)$ phase space, where it can be characterized by an emergent scale $s_0$. Within this exponential regime, $s_0$ can be approximated and, in a substantial subregion of the phase space, accurately predicted by an effective free energy $\phi$ that combines bonding, crowding, and entropic terms. This thermodynamic description can be extended to different values of $\mu$, and its applicability and accuracy naturally partition the phase space into four regimes. Our results provide a quantitative statistical-mechanical perspective on the long-standing question of self-assembly in dipolar fluids and suggest an alternative way to organize the phase space. Further studies can test how far our results and methodology can be extended to other dipolar models, such as DHS, to other topological aggregates, including rings and defect clusters, and to other physical contexts.

    The authors acknowledge the support of the United States Air Force Office of Scientific Research [grant number FA9550-23-1-0202].

\bibliography{references}

\end{document}